\documentclass[twoside]{dis08}
\usepackage[latin1]{inputenc}
\usepackage[dvips]{graphicx,epsfig,color}
\usepackage{wrapfig,rotating}
\usepackage{amssymb,amsmath,array}

\begin{document}
\title{Valence quarks and $k_T$ factorisation}

\author{Michal De\'ak, Hannes Jung, Krzysztof Kutak
\thanks{Presented by Krzysztof Kutak at the XVI Workshop on Deep Inelastic
Scattering (DIS2008), London, 7 - 11 April 2008.}
%
\vspace{.3cm}\\
%
Deutsches Elektronen-Synchrotron \\
Notkestrasse 85, 22607 - Hamburg, Germany
%
}

\maketitle

\begin{abstract}
We study in the $k_T$ factorisation framework jet production at LHC energies. 
In particular we are interested in valence quark and gluon initiated jets. 
The calculation of the relevant hard matrix element is presented. A CCFM like evolution
equation for valence quark distribution is solved and the cross section for
valence quark and gluon initiated process is calculated using Monte Carlo event
generator CASCADE.
\end{abstract}

\section{Introduction}
Collisions at  LHC will hopefully answer many so far unanswered questions
about Standard Model and beyond. Here we focus on details of jet production at LHC due to valence quark - gluon
interaction. This process is of interest since it may shed light on so called phenomenon of
saturation and it may test different factorisation schemes. This process will be measured at 
central and forward calorimeters (Castor in CMS). 

In first step we calculate hard matrix element for for $g^*q\to gq$, then we calculate the total cross section and show the differential 
cross sections as functions of
$E_T$ and $\eta$ distribution of produced jets. To calculate jet production we use CASCADE Monte Carlo event generator \cite{Jung}.

\section{Matrix element for $g^*q\to gq$}
We begin our study by calculating matrix element for $g^*q\to gq$ (this has been calculated in \cite{Ciafaloni} but in a different contest).
In our studies we assume Regge kinematics which means that we are going to work in regime where the total center of mass 
energy squared of proton-proton collision is much larger than any other scale involved in the collision. 
This practically means:
\begin{equation}\label{kinem}
s\gg \hat{s},|\hat{t}|,|\hat{u}|\nonumber
\end{equation}
where: $s=(p_A+p_B)^2, \hat s=(k+q)^2, \hat u=(k-q')^2, \hat t=(k-k')^2$ (see Figure 1).
To calculate the amplitude we have to sum expressions for diagrams
in Figure 1.
We proceed by using Sudakov decomposition of four momenta:
\vspace{-0.35cm}
$$k=x_gp_A+z_gp_B+k_T$$\\
\vspace{-1.1cm}
$$q=x_q$$\\
\vspace{-1.1cm}
$$k'=x_{g'}p_A+z_{g'}p_B+k_T'$$\\
\vspace{-1.1cm}
$$q'=z_{q'}p_A+x_{q'}p_B+q_T'$$\\
Where $x_q$ is longitudinal momentum fraction of incoming quark,
$x_g$ is a longitudinal momentum fraction of incoming gluon ($x_g\sim 10^{-5}$), $k_T$ is its transverse momentum.
$x_{q'}$ is a longitudinal momentum fraction of outgoing quark, $q_T'$ is its transverse momentum.
$x_{g'}$ is a longitudinal momentum fraction of outgoing gluon, $k_T'$ is its transverse momentum.
In calculations of matrix element the interesting kinematical regime is where the incoming
gluon carries low longitudinal momentum fraction ($x_g\sim 10^{-5}$) and we can make the eikonal approximation to the current. 
On the other hand since the incoming quark carries high momentum fraction ($x_q\sim 10^{-1}$) we take the exact expression for the current. 
The resulting formula for the matrix element squared for $g^*q\,\to
\,gq$ for unpolarized quarks and gluons, after summing and averaging over colors of final and initial state particles, reads: 
\begin{figure}[t!]
  \begin{picture}(490,170)
    \put(10, -160){
      \includegraphics{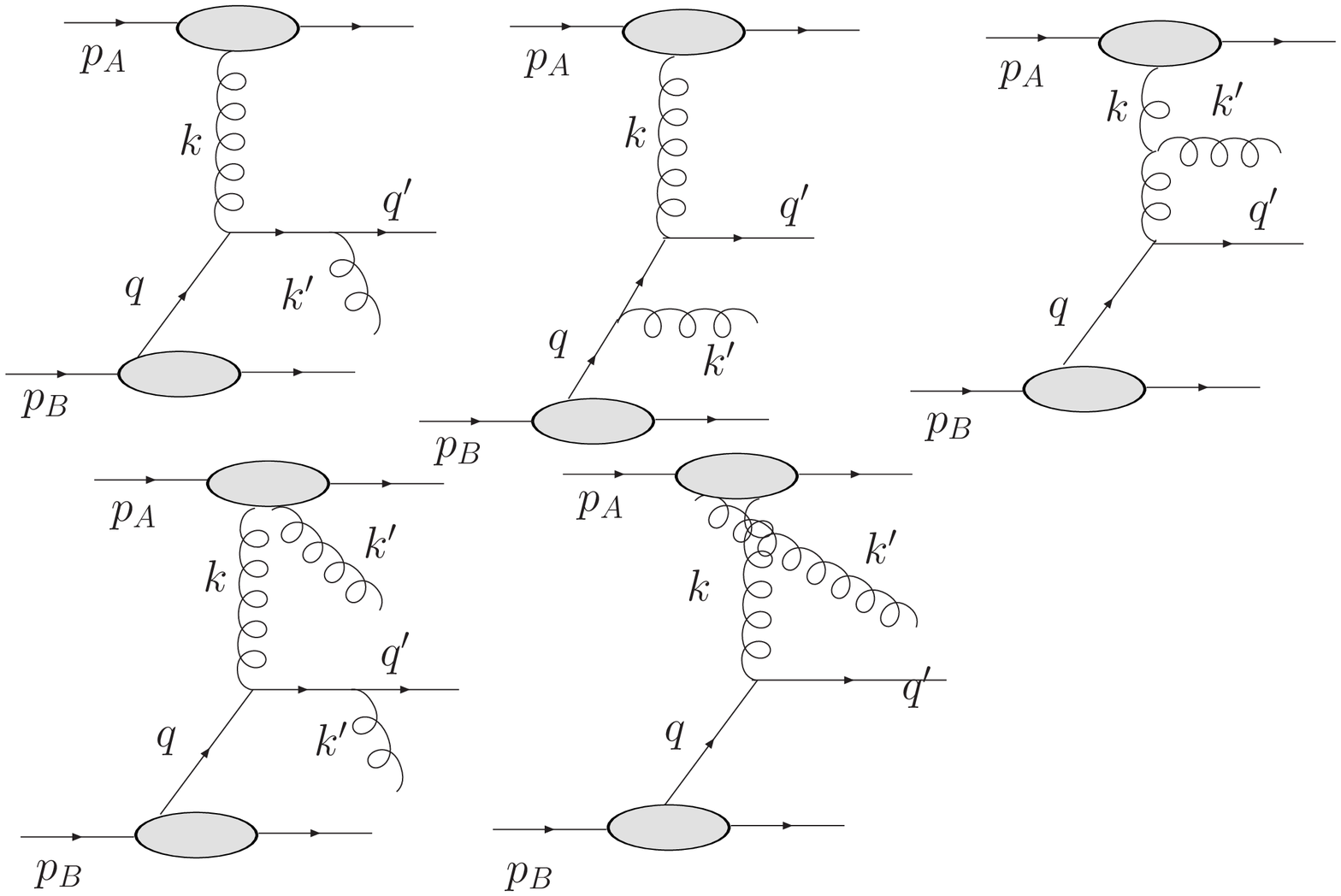}
    }

       \end{picture}
\caption{Diagrams which contribute to QCD Compton scattering in Feynman gauge. In calculation we put for proton A an auxiliary 
quark impact factor and gluon with momentum flow $k$ is of shell. Proton B is factored out and the quark with momentum $q$ is considered 
to be on-shell}
\end{figure}
\begin{equation}\label{ME2}
|\mathcal{M}|^2=\\(4\pi)^2\alpha_{S}^2\frac{x_g^2s^2(x_q^2+x_q^{\prime
2})}{18\hat{s}\hat{u}\hat{t}}\Bigg(\frac{\hat{s}(8x_q+x_q^\prime)-\hat{u}(8x_q^\prime+x_q)}
{(x_q-x_q^\prime)}-\bf{k}^2 \Bigg)\nonumber
\end{equation}
From the expression for the matrix element squared  we can see that in addition to term formally equal to
the collinear result we get a term proportional to $\bf{k}^2$.
This matrix element has interesting structure of singularities.
Singularities appear in 4 different points in the phase space when one
of the final state partons four-momentum is collinear with a four-momentum of 
a quark in the initial state (taking also the quark with the momentum $p_A$ into account).
\section{Numerical results}
The cross section in $k_T$ factorisation approach can be schematically written as:
\begin{equation}
\label{hcros}
 d\sigma(pp\to g\,q\, X ) \sim{\cal{A}}(x_g,\vec{k}_{\perp}^2,\mu^2)\otimes
M(g^*\, q \to g\,q\, X )\otimes{\cal{Q}}(x_q,\vec{q}_{\perp}^{\,\,2},\mu^2)\nonumber
\end{equation}
where $\otimes$ stands for integration over longitudinal, angular and transversal degrees of freedom, $\mu_1$, $\mu_2$ 
are the factorisation scales.
In the formula above ${\cal{A}}(x_g,\vec{k}_{\perp}^2,\mu_1^2)$ is the unintegrated, gluon density
\begin{figure}[ht!]
\centerline{\epsfig{file=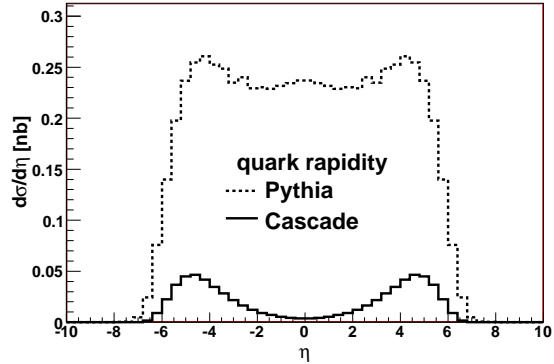,height=5.5cm}}
\caption{ Final quark peudorapidity distribution .}
\label{fig:dis}
\end{figure}
\begin{figure}[h!]
\centerline{\epsfig{file=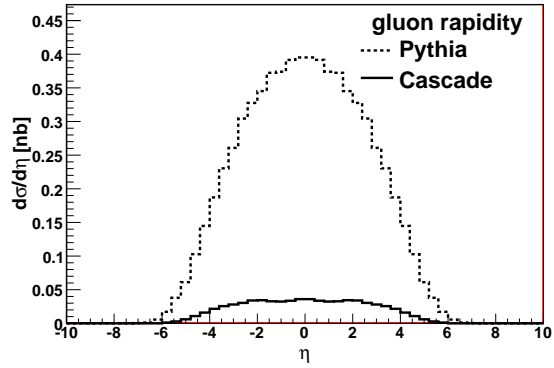,height=5.5cm}}
\caption{ Final gluon peudorapidity distribution .}
\label{fig:dis}
\end{figure}
obtained by solving the CCFM evolution equation and  ${\cal{Q}}(x_q,\vec{q}_{\perp}^{\,\,2},\mu_2^2)$
is the unintegrated valence quark density obtained from CCFM-like equation which is needed for technical reasons in
Monte Carlo simulation. The initial valence quark distribution is provided by CTEQ 6.1 set. In our calculations we use
running coupling constant and we make a $10 GeV$ cut  on transverse momenta of out going jets. This cut is motivated
by the experimental setup where jets at lower momenta are hardly measurable. This cut removes the collinear configuration 
singularities of the matrix element.
As a first observable we calculated total cross section for two jet production in considered kinematical region and we obtained $0.17mb$.  
As other observable we investigate the rapidity distributions of the produced jets.  
The results of the calculation of the rapidity distributions of outgoing partons is presented in Figure 2.
As expected, the quark initiated jet is in the forward rapidity region while the gluon initiated jet is in 
the central rapidity region. We make here a comparison to PYTHIA Monte Carlo prediction. 
\begin{figure}[t!]
\centerline{\epsfig{file=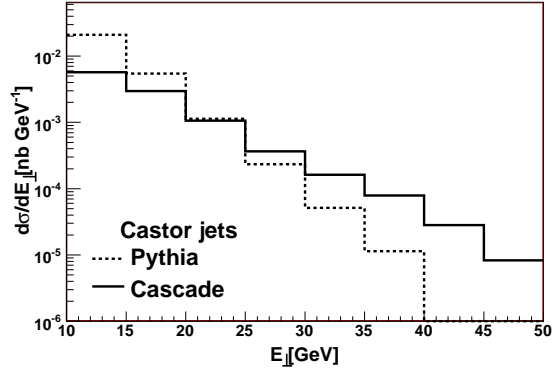,height=5.5cm}}
\caption{Transverse energy distribution of produced jet in Castor calorimeter range.}
\label{fig:dis}
\end{figure}
\begin{figure}[t!]
\centerline{\epsfig{file=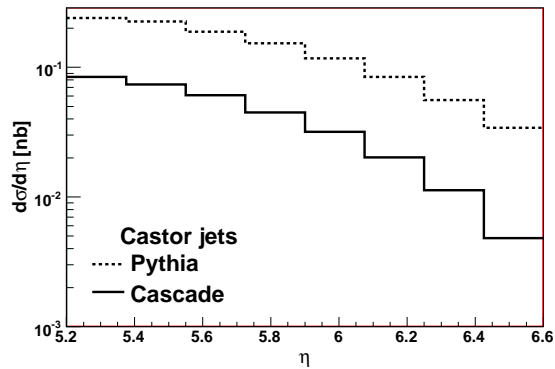,height=5.5cm}}
\caption{Pseudorapidity distribution of
produced jet in Castor calorimeter range.}
\label{fig:dis}
\end{figure}
The difference between distributions is due to missing sea quark
contribution in CASCADE.
The other interesting observable is the transverse energy $E_T$ (Figure 4) and $\eta$ distributions (Figure 5) of produced jets. The first of this distributions is of
particular importance because it shows visible difference due to different assumptions about underlying physics.
For example one sees that CASCADE including the off shell matrix
element favors a harder spectrum than PYTHIA which is based on collinear factorisation.
The $\eta$ distributions generated by CASCADE and PYTHIA differ because missing sea quarks in our approach. However, this difference
is small in region of our interest (large $E_T$). 
\section{Bibliography}
\begin{footnotesize}

\end{footnotesize}
\end{document}